\documentclass[10.5pt]{article}

\usepackage{mathptmx}
\usepackage{graphicx}
\usepackage{times}
\usepackage{amsmath, amsthm, amssymb}
\usepackage{url}
\usepackage{subfig}
\usepackage{hyperref}
\usepackage{algorithm}
\usepackage{algorithmic}

\usepackage{lscape}

\begin{document}

\begin{titlepage}

\begin{center}
{ \Large \bf
{Hierarchical sequencing of online social graphs}
}

{\large  Miroslav Andjelkovi\'c$^a$,
Bosiljka Tadi\'c$^1$
Slobodan Maleti\'c$^a$
 Milan Rajkovi\'c$^a$\\
{ $^1$Department of Theoretical Physics; Jo\v zef Stefan Institute;
 Box 3000  SI-1001 Ljubljana Slovenia;  $^a$Institute of Sciences Vin\v ca, University of Belgrade, Belgrade, Serbia}\\\hspace{1cm}}

\date{today}

\end{center}

\begin{abstract}
In online communications, patterns of conduct of individual actors and use of emotions in the process can lead to a complex social graph exhibiting multilayered structure and mesoscopic communities. Using simplicial complexes representation of graphs,  we investigate in-depth topology of online social network which is based on MySpace dialogs. The network exhibits original community structure. In addition, we simulate emotion spreading in this network that  enables to identify two emotion-propagating layers. The analysis  resulting in three structure vectors   quantifies the graph's architecture at different topology levels. Notably, structures emerging through shared links,
triangles and tetrahedral faces, frequently occur and range from tree-like to
maximal 5-cliques and their respective complexes. On the other hand, the structures which spread only negative or only positive emotion messages appear to have much simpler topology consisting of links and triangles. Furthermore, we introduce the node's structure vector which represents the number of simplices  at each topology level in which the node resides. The total number of such simplices determines what  we define as the node's topological dimension.  The presented results suggest that the node's topological dimension provides a suitable measure of the social capital which measures the agent's  ability to act as a broker in compact communities, the so called Simmelian brokerage.  We also generalize the results to a wider class of computer-generated networks.
Investigating components of the  node's vector over network layers reveals that same nodes develop different socio-emotional relations and that the influential nodes build social capital by combining their connections in different layers.
\end{abstract}
\end{titlepage}
\newpage

\section{Introduction\label{sec-intro}}
Structure of online social networks emerges via self-organizing processes of social dynamics, where the links are being established and  used for communications between individuals. The contents (information, emotion) communicated between pairs and groups of participants affects their activity patterns and thus shape the network's evolution. Two prototypal classes of online social networks can be distinguished \cite{we-entropy}: a hierarchically organized multi-layered structure that reflects the level of knowledge of the involved individuals in the chats-based systems, on one hand, and a wider class of networks with community structure, on the other.  Recently, online social networks of both types have been studied based on  high-resolution empirical data from a variety of Web portals \cite{grujic2009,mitrovic2010a,mitrovic2010b,we-MySpace11,Szell2010_,ST_Plos2013,mitrovic2010c,we-Chats1s,we-Chats-conference,we-Chats-chapter,we_ABM_blognets}.

In social sciences, the \textit{value of social networking} can  be quantified. Specifically, the \textit{social capital}, as an essential element of functioning social networks, measures how social relationships create a competitive advantage of certain individuals at the expense of others \cite{taube2004}. For instance, the Simmelian ties and the related brokerage roles measure how an individual can act as a broker (transmitting actor) in a dense social community \cite{panzarasa_brokerage}. The availability of massive empirical data of online social interactions and their dynamics that completes the information for the study of social graphs offers the possibility to investigate the mechanisms of social capital build-up by different actors. In view of ever-evolving social graphs, intricate communication processes result in the social-value  relationships within communities where the social layers may play a role. Analysis of such social structures on multilayered networks requires new mathematical approach, capable of taking into account the inner (in depth) structure of each society and the role of different nodes in it.

In recent research, efforts have been made on determining the network complexity metric that permits to successfully distinguish between critical and redundant nodes \cite{nacher2014}, discover the active cores of the network as compared to the network's periphery \cite{csermely2013,we-Chats-conference} and quantify the network's multiplexity\cite{szell2010,gomez2012,lee2012,cardillo2013,ST_Plos2013,we-Chats-chapter,estrada2013social} and the role of higher-order structures in the network dynamics \cite{maletic2014,milan2012,tadic2007}. In this respect, several approaches have focused on introducing suitable graph-theoretic vectors that can be  defined on local graphlets \cite{natasa_graphlets}, the network's feature vector \cite{costa_vector} or the graph's eigenvalue spectrum \cite{estrada_4classes}.
These approaches proved very useful in the study of biological systems, for example, in alignment of protein networks \cite{natasa_proteins2014} and uncovering network function in cancer-related processes \cite{natasa_cancer2008}. Similarly, characterizing topology of molecular graphs \cite{estrada_molecular} as well as  determining modules in socio-technological networks by
eigenvectors localization \cite{mitrovic2009} have been successful.

In this work, we exploit the topological concept of a \textit{simplex}, structure that extends beyond the nodes and links, i.e., a polyhedron of possibly high dimension, and their aggregates or \textit{simplicial complexes}. We  investigate in-depth topology of online social networks and explore the role of nodes in layers and communities. The concept of simplicial complexes of graphs \cite{SimplComplexesOfGraphs,milan_ICCS2008} allows precise definition, using topological, algebraic and combinatorial tools, of the node's natural surroundings in the network. Consequently, it allows to determine the node's social capital such as the Simmelian brokerage in social environments.

The considered network  in this paper is constructed from the original data collected from \texttt{MySpace} social network as described in Ref.\ \cite{we-MySpace11}. Typically, in  online social networks, such as \texttt{MySpace} and \texttt{Facebook}, certain  kind of  social graphs exist a priori. However, the use of connections over time as well as the dominance of positive emotions in the texts of messages \cite{we-MySpace11} reveal the dynamical structure that is different from the conventional social networks.

In general, network layers appear due to different types of relationships
among nodes \cite{multiplexity_review2014,we-Chats-conference,we-Chats-chapter}. In \texttt{MySpace} and \texttt{Facebook} social networks, where text messages of mixed information contents are communicated, the emotion contained in these words can be inferred \cite{gpalt,we-MySpace11}. This  fact offers the possibility to identify the network's layered structure in a unique manner. Specifically, one can define layers that propagate emotions with positive or negative valence (attractiveness and aversiveness). Quantitative study of emotions, based on Russell's model \cite{russell1980}, and the social dimension of emotional interactions are the subject of an intensive research in recent years (see the summary of the Cyberemotions project \cite{CYB}).  In this context,   the  dynamics of emotion spreading on networks has been investigated by an agent-based model \cite{we-MySpaceABM, we-Cybbook}. In this paper, we employ this model to generate network layers. The agents are interacting along the social links of \texttt{MySpace} network influencing each-other's emotional state by exchanging messages. The identity of each message in the simulations is known, its time of inception,  source and recipient nodes as well as its emotional content that matches the current emotional state of the agent who created it. Consequently, the links carrying  negative/positive emotion messages can be distinguished; together with the involved nodes these links define the corresponding network layer.

The paper is organized as follows. The standard structure of the network, as well as the dynamics leading to the emotion-propagating layers, is briefly described in Section\ \ref{sec-network}.
 The basic definitions and the method are described in \ref{sec-simplices}. The analysis of simplices and their complexes in the communities, section\ \ref{sec-topology}, and in the emotion-propagation layers of the network, section\ \ref{sec-layers}, is presented.  Section\ \ref{sec-brokerage} deals with the functional relationship between Simmelian brokerage role of a node and its topological dimension. Section\ \ref{sec-conclusions} contains a brief summary of the results and conclusions.

\section{Simplices: Beyond standard structure of social graphs\label{sec-communities}}

\subsection{Dynamic architecture of online social networks\label{sec-network}
}
In this work we use the online social network from the  empirical data of Ref.\ \cite{we-MySpace11}, consisting of the links in \texttt{MySpace} along which the messages were exchanged within 2-months time window. The original data and the network mapping are described in \cite{we-MySpace11} together with  the study of the dynamics of emotions detected in the data.  The structure of these networks has been defined in terms of  several measures. In particular, the degree- and strength- distributions, link correlations---disassortativity, reciprocity, path lengths, clustering, community structure, and testing the weak-tie hypothesis have been determined in Ref.\ \cite{we-MySpace11} and for the purpose of this work, this study will be termed  \textit{standard analysis of a social graph}.  These results revealed that the \texttt{MySpace} dialogs graph exhibits the characteristic social community structure. A similar conclusion was derived in Ref.\ \cite{facebook2011} considering the static structure of the \texttt{Facebook} graph. However, the analysis in \cite{we-MySpace11} uncovers that several other topology measures, notably, disassortativity, non-reciprocal in- and out-linking, the role of weak-ties, are different from the features  often found in social networks. Moreover, they share a high similarity with the networks of online games studied recently in \cite{Szell2010_}. Together with some other findings \cite{myspace2007,facebook2012}, the observations in \cite{we-MySpace11} suggest that the dynamical organization  of online social networks can be  strikingly different from  that of the conventional social graphs. The origin of such structure can be traced in the dynamics of exchanged messages with their emotional contents \cite{gpalt,we-MySpace11,dodds2011,we-MySpaceABM}.  Hence, the patterns of conduct and  the role of individuals in these processes \cite{panzarasa2009,ST_Plos2013,ST_triadic} are crucial for the emergence of macro-social temporal structures. Therefore, an in-depth topology beyond the standard analysis of the emergent social networks can reveal some of the working mechanisms at the local level.
In the remaining parts of this work, we extend the analysis of these networks by means of simplicial complexes.  For this purpose, the original network
which consists of 32000 nodes is reduced to 3321 nodes by removing nodes with
a low connectivity \cite{we-MySpace11}.  The dynamics of emotional messages
is simulated using the agent-based model represented in \cite{we-Cybbook}.

\subsection{Simplices and simplicial complexes of graphs\label{sec-simplices}
}

In contrast to the standard graph theoretical approach which reveals
connectivities, a topological framework \cite{SimplComplexesOfGraphs}
provides information about the structure and patterns of connectivities and enables the multifaceted approach including topological, algebraic and combinatorial techniques. A
basic element of the topological approach is a simplex whose spatial
representation is a polyhedron. The polyhedra of various dimensions (i.e.
the number of vertices and faces) may be connected to each other forming a
polyhedral complex, the spatial embodiment of a simplicial complex.
In a more formal approach, any subset of a set of vertices $V=$ $\{v_{\alpha
_{0}},v_{\alpha _{1}},...,v_{\alpha _{n}}\}$ determines an $\ n$-$simplex$
denoted by $\left\langle v_{\alpha _{0}},v_{\alpha _{1}},...,v_{\alpha
_{n}}\right\rangle $ where $n$ is the dimension of the simplex. A $q$%
-simplex $\sigma _{q}$ is a $q$-face of an $n$-simplex $\sigma _{n}$,
denoted by $\sigma _{q}\lesssim \sigma _{n}$, if every vertex of $\sigma
_{q} $ is also a vertex of $\sigma _{n}.$ The union of faces of a simplex $%
\sigma _{n}$ is a boundary of $\sigma _{n}.$ A simplicial complex represents
a collection of simplices and its dimension is the largest dimension of its
simplices. More formally, a simplicial complex $K$ on a finite set $%
V=\{v_{1},...,v_{n}\}$ of vertices is a nonempty subset of the power set of $%
V$, so that the simplicial complex $K$ is closed under the formation of
subsets. Hence, if $\sigma \in K$ and $\rho $.$\in $ $\sigma ,$ then $\rho $.
$\in K.$ An example of a simplex of dimension zero is a point. A simplex of
dimension one is a line; two-dimensional simplex is a triangle,
three-dimensional simplex is tetrahedron and so on.

Simplicial complexes may be constructed directly from the available data or
they may be formed from the undirected or directed graphs (digraphs) in
several different ways. Here we only consider two of them: the neighborhood
complex \cite{neighborcomplexes} and the clique complex \cite{cliquecomplexes}. The neighborhood complex \emph{N}$(G)$ is
constructed from the graph $G$, with vertices $\{v_{1},...,v_{n}\}$ in such
a way that for each vertex $v$ of $G$ there is a simplex containing the
vertex $v$, along with all vertices $w$ connected to it and the
corresponding faces. The neighborhood complex is obtained by including all
faces of those simplices and in terms of matrix representation, the
incidence matrix is obtained from the adjacency matrix of $G$ by increasing
all diagonal entries by $1$.

The vertices of the clique complex $C(G)$ are the same as the verties of $G$
with the maximal complete subgraphs (cliques) as simplices so that it is
essentially the complete subgraph complex. An example of the clique complex
construction is presented in Fig. $1$.

\begin{figure}[hb]
  \centering
  \resizebox{14.8pc}{!}{\includegraphics{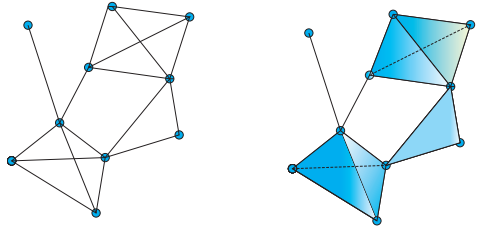}}\\
  \caption{Construction of the clique complex (right) from the graph
(left)}
\end{figure}

These two methods are not the only ones which may be used for constructing
simplicial complexes from graphs. Actually, any property of the graph $G$
that is preserved under deletion of vertices or edges may be used for
construction purposes. A detailed account of the methods for obtaining
simplicial complexes from graphs, among many other issues related to the
relationship between graphs and simplicial complexes, may be found in \cite{SimplComplexesOfGraphs}.

Initially all maximal cliques (MC) were found using the Bron-Kerbosch
algorithm \cite{Bron}. The resulting MC matrix contains full information
about cliques, the clique's ID, size and IDs of nodes that are involved in
it; from this the connection between different simplices can be deduced \cite{Bron}, \cite{milan_ICCS2008}.

As mentioned earlier, the power of the topological approach is based on the
fact that simplicial complexes may be considered from three different aspect:
(1) a combinatorial model of a topological space; 2) a combinatorial object;
3) an algebraic model. Hence, different measures and invariants may be
associated to the simplicial complex based on an appropriate perspective,
and each one of them provides new information about the graph or network
from which the simplicial complex was constructed. The first is the
dimension of the simplicial complex corresponding to the maximal simplex
dimension, $K$. \ From the combinatorial aspect three structure vectors of
the simplicial complex are defined \cite{milan_ICCS2008}:

\begin{itemize}
\item the \textit{first structure vector} $Q$: The $q^{t}h$ entry of the so
called $Q-vector$ of length $K+1$(or \textit{first structure vector} \cite{maletic2014}), denoted by $Q_{q}$ is equal to the number of $q$-connectivity
classes. This vector provides information on the number of connected
components at each level of connectivity with initial level being equal to
the dimension of the complex:
\begin{equation}
\mathbf{Q}=\{Q_{q=K}Q_{q=K-1}\dots Q_{q=1}Q_{q=0}\}\ ;  \label{eq-1strvector}
\end{equation}

\item the \textit{second structure vector} $N_{s}$ is an integer vector with
$dim(K)+1$ components
\begin{equation}
\mathbf{N_{s}}=\{n_{q=K}n_{q=K-1}\dots n_{q=1}n_{q=0}\}\ ;
\label{eq-2strvector}
\end{equation}%
where the $q-th$ entry, $n_{q}$, is equal to the number of simplices with
dimension larger or equal to $q$, that is, it is equal to the number of
simplices at the q-level.

\item the \textit{third structure vector} $\hat{Q}$ is the global
characteristic of the simplcial complex which measures the degree of
connectedness on a $q$-level. In other words, it measures the number of $q$-connected components per number of simplices whose components $\hat{Q}_{q}$
are defined as \cite{SimplComplexesOfGraphs}:
\begin{equation}
\hat{Q}_{q}=1-\frac{Q_{q}}{n_{q}}\ ;  \label{eq-3strvector}
\end{equation}%
where $Q_{q}$ is q-th entry of the first structure vector, and $n_{q}$ is
q-th entry of the second structure vector.
In addition to these structure vectors that describe the network as a whole or its large parts, in the following we will introduce the vector associated to each node, the node's $Q$-vector,  in order to characterize neighborhood of the node in the network.
\end{itemize}
\begin{figure}[!htb]
  \centering
\begin{tabular}{ccc}
\resizebox{12.8pc}{!}{\includegraphics{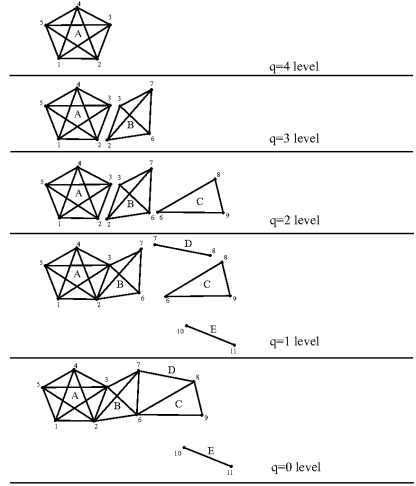}}\\
\end{tabular}
  \caption{Q-vector disclosing connectivity of the complex at various $q$-levels. The underlying graph is at the $q=0$ level.}
\end{figure}

\subsection{Simplicial complex analysis of the social network with
communities\label{sec-topology}}

\begin{figure}[h]
\centering
\begin{tabular}{cc}
{\large (a)} & {\large (b)} \\
\resizebox{12.8pc}{!}{\includegraphics{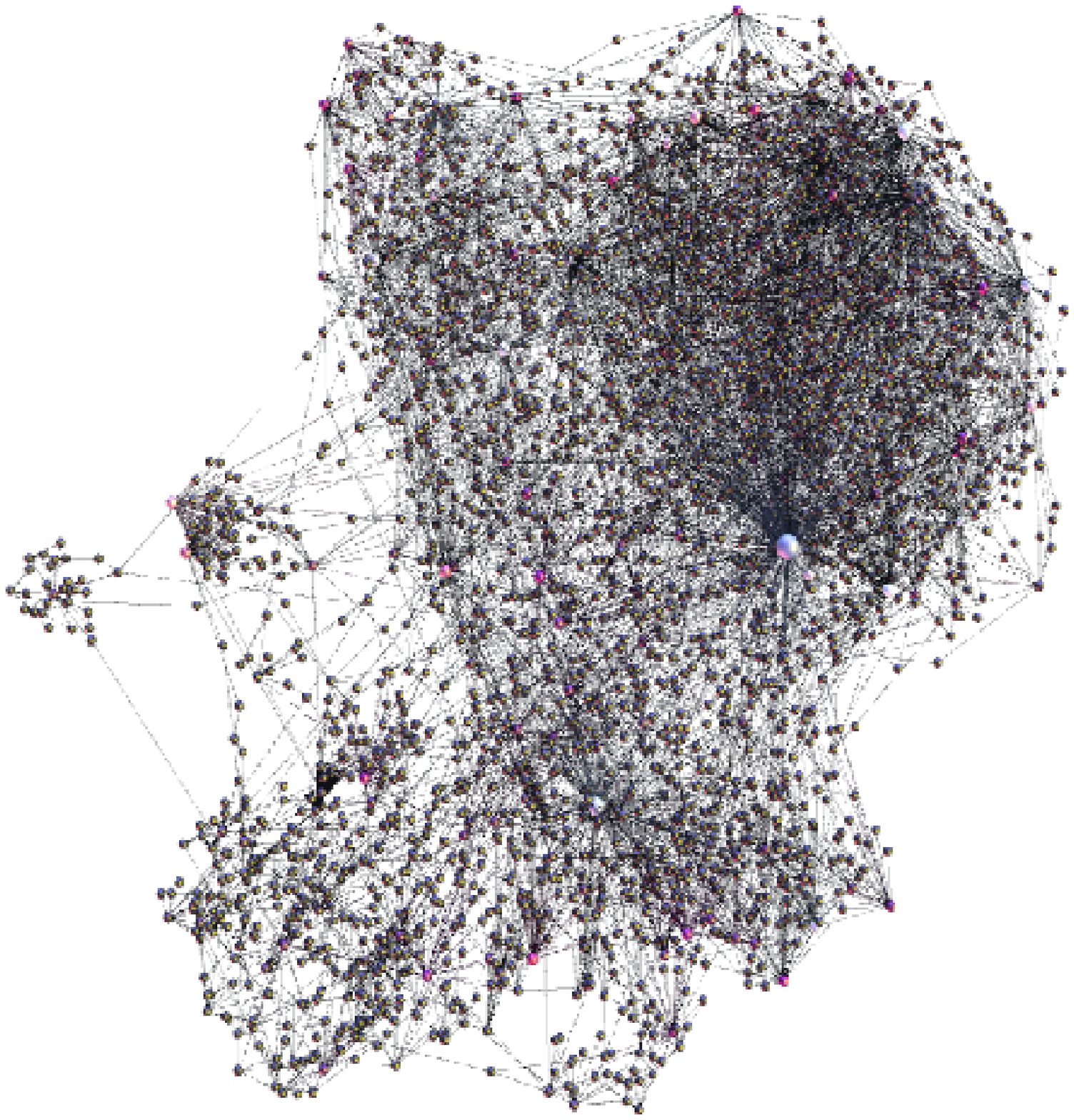}} &
\resizebox{12.8pc}{!}{\includegraphics{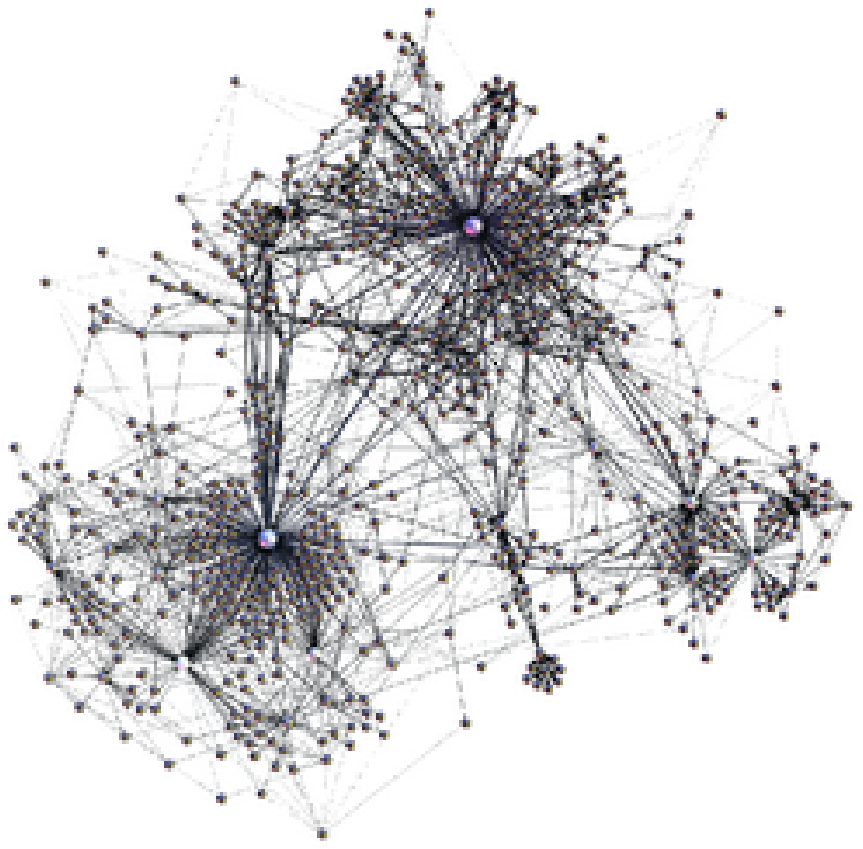}} \\
\resizebox{12.8pc}{!}{\includegraphics{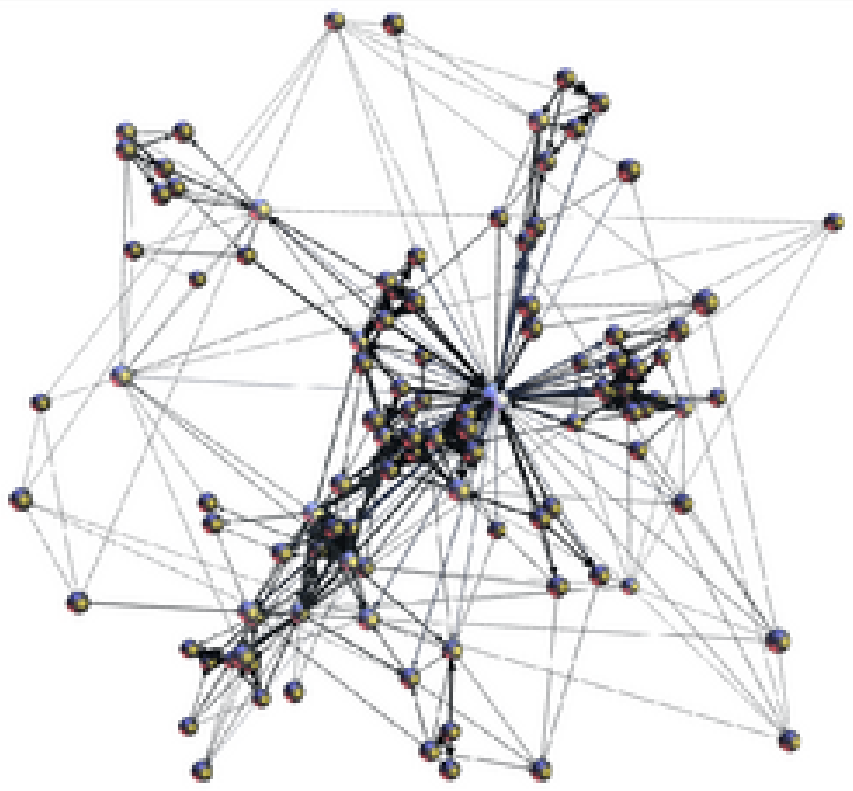}} &
\resizebox{12.8pc}{!}{\includegraphics{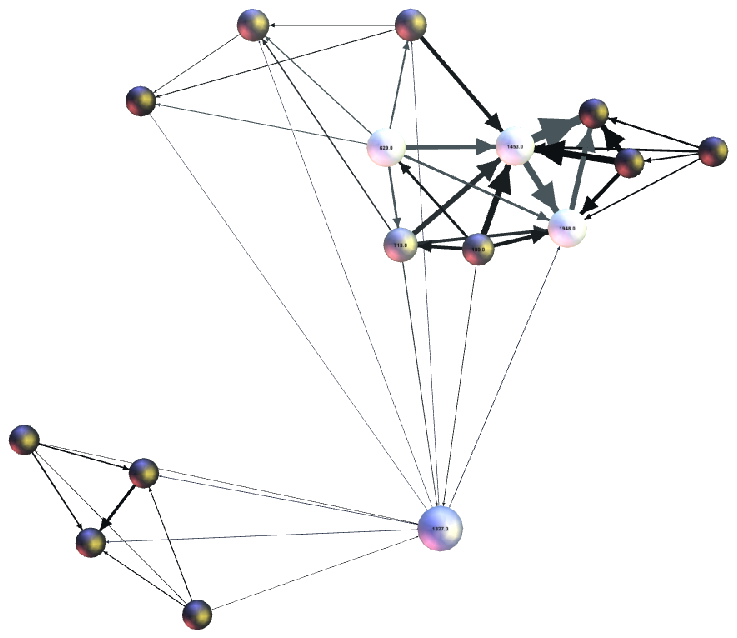}} \\
{\large (c)} & {\large (d)} \\
\end{tabular}%
\caption{Online social network, constructed from \texttt{MySpace} dialogs, after removal of leaves (a); Structure of the largest social community
in that network at different topology layers corresponding to dimensions $q$
= 2, 3 and 4 (b,c, and d, respectively).}
\label{fig-osn_q}
\end{figure}

In this section, we analyze a reduced system containing $N_{U}=3321$ nodes and no leaves. Focusing on the large community in this network, we perform the analysis on simplices and determine the network's MC matrix. It appears that the maximal clique simplices (in further text for brevity we use term clique) in the network are 5-cliques, i.e. in the present  notation, $q_{max}=4$. The analyzed structures at different $q$-levels are shown in Fig.\ \ref{fig-osn_q}.

As $q$ increases, the number of nodes that participate in the higher topological structures is decreasing. Considering the largest community in the network shown in  Fig.\ \ref{fig-osn_q}a, we compute its structures at higher topological layers; they are shown in Fig.\ \ref{fig-osn_q}b,c,d corresponding to the level of triangles, $q=2$, 4-cliques, $q=3$, and at the level of 5-cliques, $q=4$. Summary of the network's structure vectors will be given in Section\ \ref{sec-brokerage} in connection with the discussion of brokerage roles of nodes.

Here,  focus is on the role of individual nodes. Therefore, we introduce the "node's $Q$-vector", $Q^{i}$, that describes the node's $i$ environment in the network. \\
\textit{Definition of the Node's Q-vector.} With $q_{max}+1$ as the dimension of the maximal clique in the network, the \textit{node's Q-vector} is a $q_{max}+1$--dimensional vector  associated with  each node
\begin{equation}
Q^{i}=\{Q_{qmax}^{i},Q_{qmax-1}^{i},...,Q_{1}^{i},Q_{0}^{i}\}\ ,
\label{eq-Qi}
\end{equation}%
whose components $\{Q_{q}^{i}\}$, $q=q_{max},q_{max}-1,\cdots 1,0$ describe the number of $q$-dimensional simplices in which the node $i$ participates. We define the \textit{topological dimension} of the node $i$ as the number of all simplices in which the node $i$ participates, i.e., $dimQ^{i}= \sum_{q=0}^{q_{max}}Q_{q}^{i}$. The term "dimension" is motivated in view of the conjugate simplicial complex constructed from the network, where nodes become simplices and simplices become nodes. Therefore, $dimQ^{i}$ corresponds to the dimension of the conjugate simplex \cite{milan_ICCS2008}. To illustrate the meaning of these components consider the vector  $Q^i=\{2,0,3,2,0\}$ in the network whose maximal clique dimension is 5. In this case, the value of the component $Q_{0}=0$ suggests that the node is not isolated. Further, $Q_{1}=2$ means that among all the nodes connected to the node $i$, only two are not a part of any higher-order clique. $Q_{2}=3$ indicates that there are three triangles attached to $i$ that are not faces of higher-order cliques; while $Q_{4}=1$ and $Q_{3}=0$ indicate that the node $i$ belongs to one 5-clique, but there are no 4-cliques attached to it.

\begin{figure}[tbh]
\centering
\begin{tabular}{ccc}
\resizebox{24pc}{!}{\includegraphics{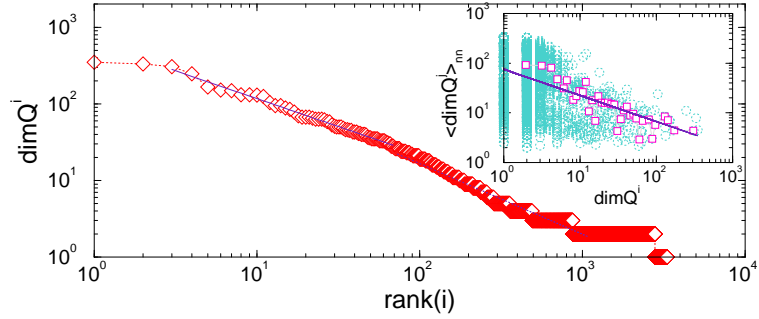}}\\
\end{tabular}%
\caption{Ranking plot (Zipf's law) of topological dimensions $dimQ^{i}$ of
all nodes in the network, main panel. Inset: Average topological dimension
of the node's $i$ nearst neighbors plotted against the node's $i$ dimension
exhibiting disassortative behavior (slope of the fit line is $\protect\mu
=-0.527\pm 0.026$). }
\label{fig-dimQ-Zipf}
\end{figure}

The components $\{Q_{q}^{i}\}$ for each node in the analyzed network can be computed from the MC matrix. Here we determine the components of each node of the network in Fig.\ \ref{fig-osn_q}a. Sorting the nodes according to their topological dimensions $dimQ^{i}$, one  can identify the \textquotedblleft influential\textquotedblright\ nodes in the network's community or a layer.  Fig.\ \ref{fig-dimQ-Zipf}, shows the distribution of the node's dimension for all nodes in the network as a function of the node's rank. Notice that the topological dimensions exhibit a broad distribution (Zipf's law) with two rather than a single slope. Such situation often appears in the evolving complex systems \cite{katz2006,we-genes}.
Recently, the origin of two slopes in the Zipf's law has been discussed  \cite{al-zipf} in connection with the scaling and innovation in the use of  words in the written text of an increasing length.
In the present  case, the appearance of new topological forms in the dialogs-based network is related with the activity patterns of users (nodes in the network). According to the analysis in \cite{we-MySpace11}, three distinct groups of users can be distinguished considering the number of their actions in relation with the interactivity times. Consequently, very active nodes may build a larger environment resulting in a higher topological dimension.  In Fig.\ \ref{fig-dimQ-Zipf}, the nodes topological dimensions obey a power-law distribution with the slope $\gamma =0.72\pm 0.05$ up to the rank 70 (corresponding to the density function power-law with the exponent $\tau =1/\gamma +1\approx 2.37)$.  In contrast, the nodes of lower dimensions (higher ranking orders) show the slope close to one, $\gamma =0.92\pm 0.06$). Temporal appearance of new topology forms will not be considered in this paper. In the following, we study correlations among topological dimensions of the connected pairs of nodes.

In an analogy with standard assortativity measure in social networks \cite{newman2003}, we plot the node's topological dimension against the average topological dimension of its neighbors. The results, shown in the inset of Fig.\ \ref{fig-dimQ-Zipf}, indicate that at the level of triangles and cliques of higher dimension, the graph exhibits disassortativity. The general trend of all points can be approximated with the function $\langle dimQ^{j}\rangle _{nn}\sim (dimQ^{i})^{-0.52}$. This means that gradually fewer number of nodes with high dimension are connecting between structures of a lesser complexity. These findings complement the disassortativity results for the same network found in \cite{we-MySpace11} at the level of links (i.e., including leaves) for varied combinations of the directed link orientations.

\begin{figure}[tbh]
\begin{tabular}{ccc}
{\large (a)} & {\large (b)}  \\
\resizebox{14pc}{!}{\includegraphics{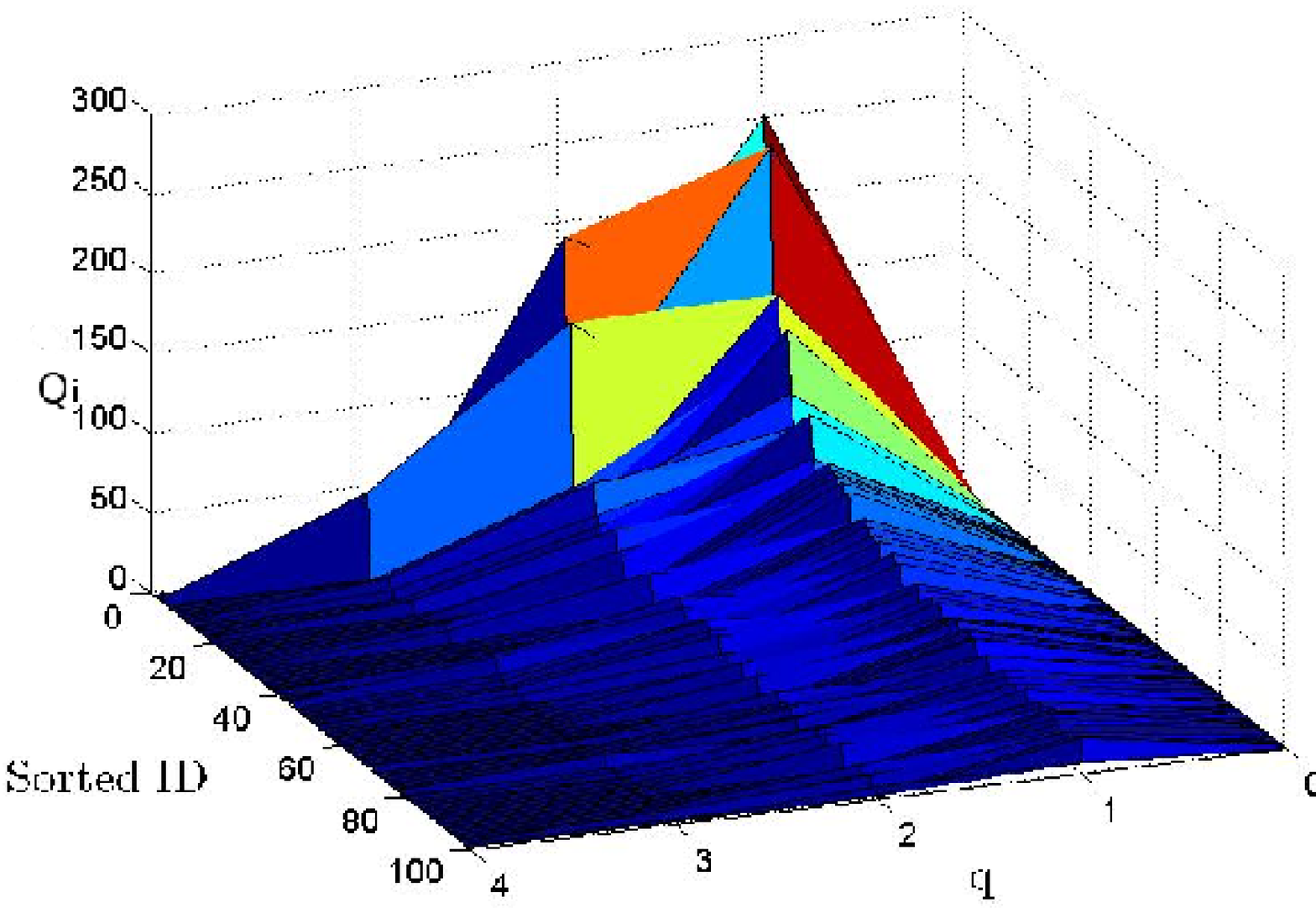}}&
\resizebox{14pc}{!}{\includegraphics{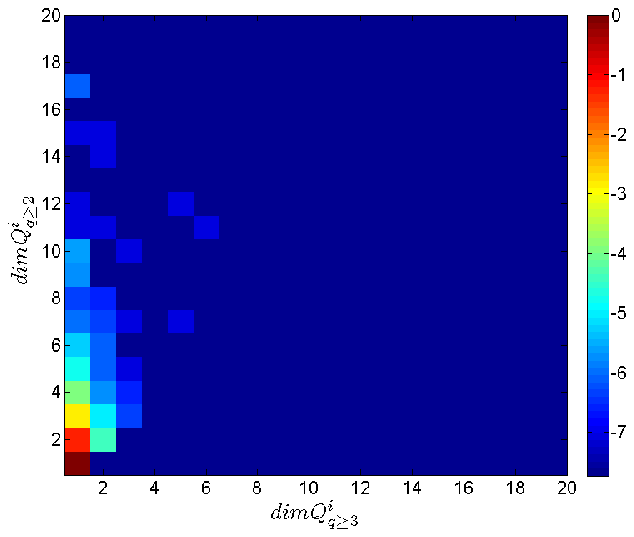}}\\
\end{tabular}%
\caption{ (a) Components of the topology vector $Q_{q}^{i}$ of the first 100
nodes plotted against node's index $i$ and simplex dimension $q$. Ranking
order of nodes according to the topological dimansion $dimQ^{i}$ applies.
(b) Scatterplot of the node's $i$ topological dimension $dimQ_{q\geq 2}^{i}$
against its dimension at higher level of clicques, $dimQ_{q\geq 3}^{i}$. }
\label{fig-qq-3D}
\end{figure}

Fine structure of a node $Q$-vector allows for further differentiation between nodes and their roles in the network. A 3-dimensional plot in Fig.\ \ref{fig-qq-3D}a shows the components $Q_{q}^{i}$ of the first 100 nodes  as a function of the node's dimension $dimQ^{i}$ and the dimension of the corresponding simplices $q$. It also shows the number of triangles ($q=2$), tetrahedra ($q=3$) and 5-cliques to which the leading nodes belong.
In Fig.\ \ref{fig-qq-3D}b a scattered plot of the node's dimension computed
from the topology layer $q\geq 2$ is shown against $q\geq 3$. Different
colors reflect the number of nodes corresponding to a given combination in
the plot. For each node a nonzero value in this plot $(dimQ_{q\geq
3}^{i},dimQ_{q\geq 2}^{i})$ indicates the number of triangles $n_{\triangle
}^{i}=dimQ_{q\geq 2}^{i}-dimQ_{q\geq 3}^{i}$ surrounding that node that are not
shared faces of higher-order structure. Obviously, the number of nodes which
participate in higher-order structures (of higher dimension) is steadily
decreasing. Thus, compared with a standard clustering coefficient, the
in-depth-topology analysis with simplicial complexes differentiates between
types of structures in which the node resides.

The nodes that build up these higher-order structures can be clearly distinguished owing to the unique ID assigned to every node. As it will be presented in Sec.\ \ref{sec-brokerage}, these nodes often have a considerable social capital. An example is the node with ID=1372, connecting three 5-cliques in Fig.\ \ref{fig-osn_q}d.  The composition of its components in the entire network is given in Table\ \ref{tab-2nodes}.  Apart from three 5-cliques, this node joins 53 different 4-cliques at lower $q$-levels, 144 triangles and 46 sigle-link simplices. Another type of the node with a large social capital is the central node in a predominantly star-like structure; an example is the node with ID=2238, also given in Table\ \ref{tab-2nodes}.
The changed roles of these and other nodes on emotion-propagating layers of the network are studied in the subsequent sections.

\section{Structure Vectors of Emotion-Propagating Network Layers\label{sec-layers}}

\begin{figure}[!htb]
\begin{tabular}{ccc}
{\large (a)} & {\large (b)}   \\
\resizebox{14pc}{!}{\includegraphics{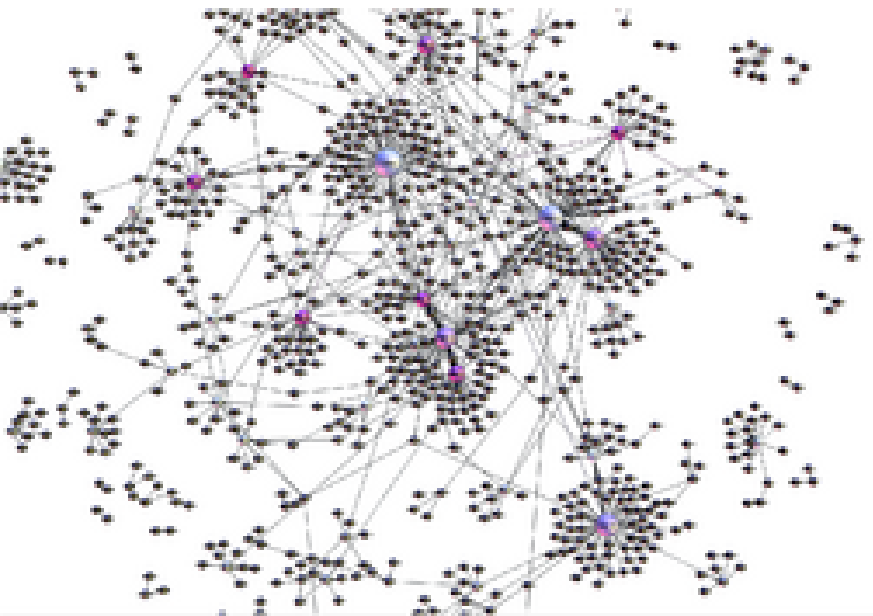}} &
\resizebox{14pc}{!}{\includegraphics{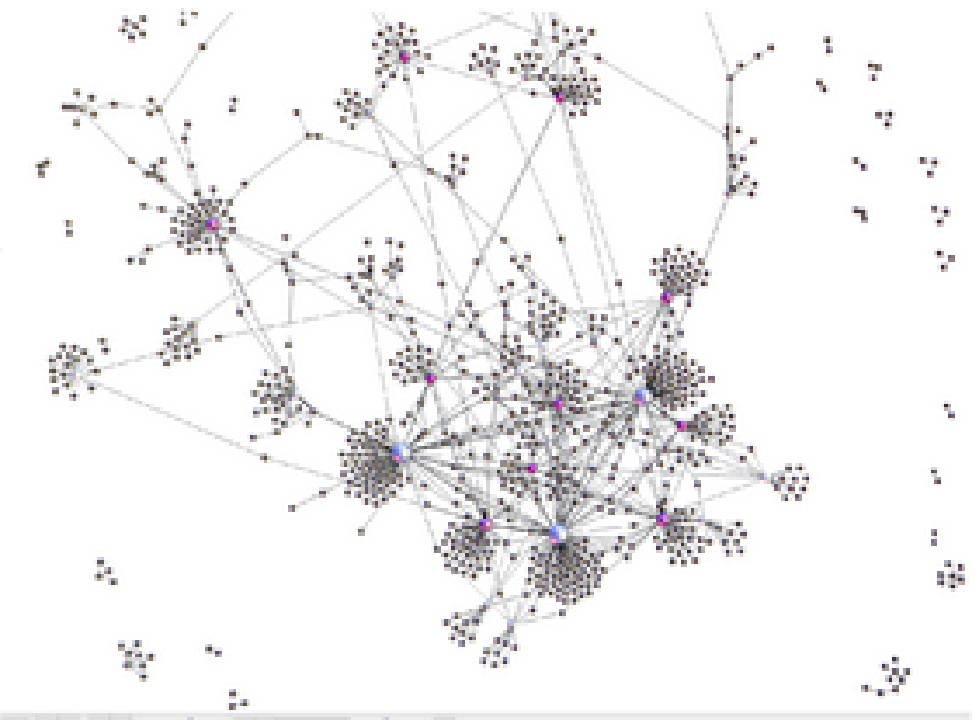}} \\
\end{tabular}
\caption{Counter-emotion layers NP (left) and PN (right) of the online social network from Fig.\ \ref{fig-osn_q}a.}
\label{fig-nets_couterlayers}
\end{figure}

As discussed in section\ \ref{sec-network}, the emotion-propagation dynamics involves different types of contacts among individuals in online social network. The diversity of the emotional content of communicated messages, described by two variables--emotional arousal and valence, enables to identify different network layers corresponding to a particular type of emotional content. Consequently, two network layers are recognized. The positive layer consists of the links along which messages with a positive emotion valence were communicated up to a given instance of time, and the negative  layer with links carrying messages with negative valence. In order to enhance the difference between these layers, using the agent-based model of Ref.\ \cite{we-MySpaceABM}, we numerically simulated two situations. In one,  the external input noise has prevailing positive emotion "astonished" and, in  the other,  the negative emotion "ashamed". As it was shown in \cite{we-MySpaceABM,we-ABMrobots,we-ABM_bots2}, in such situations the temporal correlations of message streams, that marks  the emotion propagation dynamics, lead to collective emotion states in networks. Here,  the positive/negative valence of the input emotion eventually prevails. In this way, in each case one can distinguish a dominant layer that diffuses the ``winning'' emotion from the counter-emotion layer. These are  named PP (positive-positive) and NN (negative-negative) as dominant layers in positive and negative input, respectively; counter-emotion layers are NP and PN.
Here NP designates a layer with links propagating positive emotion in the case when the majority of messages in the network are negative (following negative emotion input). While PN denotes a layer with the negative emotion links when the prevailing emotion is  positive. Note that, by definition, the same nodes (and sometimes overlapping links between them) can belong to both layers.
The focus is on in-depth-topology analysis of these layers and in quantifying the roles of relevant nodes residing in each one of them. The structure of connections in counter-emotion layers for the two cases of emotion-propagating dynamics is shown in Fig.\ \ref{fig-nets_couterlayers}. The results of topology analysis of all emotion-propagating layers are summarized in Table\ \ref{tab-layers_q}.

\begin{table}[h]
\caption{Layers in the OSN propagating negative and positive emotion
messages when the majority of messages are of negative emotional content (NN
and NP) and when the majority of messages in the network are of positive
emotional content (PN and PP). Compared with other structure vectors for the
whole network, given in Table\ \protect\ref{tab-nets},
in the emotion-propagating layers 5-cliques ($q=4$ components of the
structure vectors) are absent; in counter-emotion layers 4-cliques also do not appear.}
\label{tab-layers_q}\centering
\begin{tabular}{|c|ccc|ccc|}
\hline
& \multicolumn{3}{c|}{$NN_{w>30}$} & \multicolumn{3}{c|}{$NP$}  \\ \hline
$q$ & $Q$ & $N$ & $\hat{Q}$ & {$Q$} & $N$ & $\hat{Q}$  \\ \hline
4 & - & - & - & - & - & -   \\
3 & 4 & 4 & 0 & - & - & - \\
2 & 198 & 198 & 0 & 12 & 12 & 0   \\
1 & 6242 & 6353 & 0.017 & 1134 & 1138 & 0.003  \\
0 & 1 & 6353 & 0.999 & 2310 & 3404 & 0.321  \\
\hline
& \multicolumn{3}{c|}{$PN$} & \multicolumn{3}{c|}{$PP_{w>30}$}  \\ \hline
$q$ & $Q$ & $N$ & $\hat{Q}$ & $Q$ & $N$ & $\hat{Q}$   \\ \hline
4 & - & - & - & - & - & -   \\
3 & - & - & - & 5 & 5 & 0   \\
2 & 11 & 11 & 0 & 204 & 205 & 0.005   \\
1 & 1069 & 1074 & 0.005 & 6228 & 6287 & 0.009   \\
0 & 2360 & 3390 & 0.304 & 4 & 6352 & 0.999  \\ \hline
\end{tabular}%
\end{table}

As may be noticed in Fig.\ \ref{fig-nets_couterlayers} and Table\ \ref{tab-layers_q}, the counter-emotion layers consist of a reduced number of connections as compared to the dominant-emotion layer and the whole system. Furthermore, the structure of connections is different, depending on the dominant emotion, and much simpler than in the dominant-emotion layer. The cliques of dimension 4 and 5 existing in the simplicial complex of the network are absent in the counter-emotion layers. The highest cliques are triangles ($q=2$ components of the structure vectors). Comparison of the first and the second structure vectors of NP and PN layers shows that only  few triangles, 4 in the case of NP an 5 in PN layer, join the other simplices at the $q=1$ level. In both cases, a large number of disconnected components remains at the $q=0$ level, in contrast to the dominant layers. The dominant-emotion layers (PP and NN) are richer, although the number of 4-cliques is much smaller than in the entire network, and the 5-cliques are still absent. For comparison, the structure vectors of the entire network are given in Table\ \ref{tab-nets}  when all links are counted irrespectively of the type of emotion diffusing along them.

\begin{figure}[tbh]
\begin{tabular}{ccc}
\centering \resizebox{14pc}{!}{\includegraphics{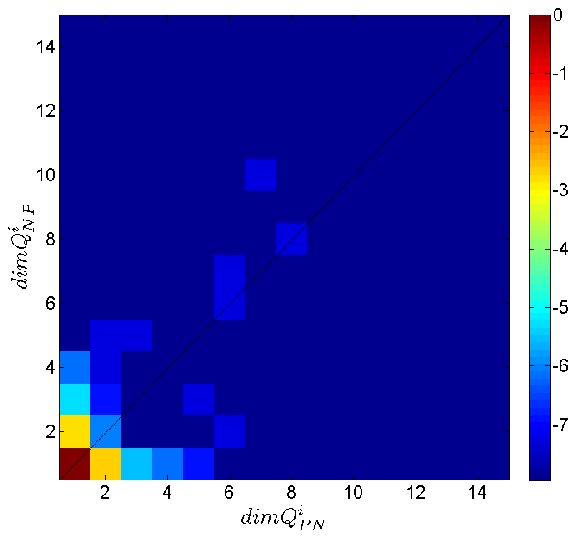}} &
\resizebox{16pc}{!}{\includegraphics{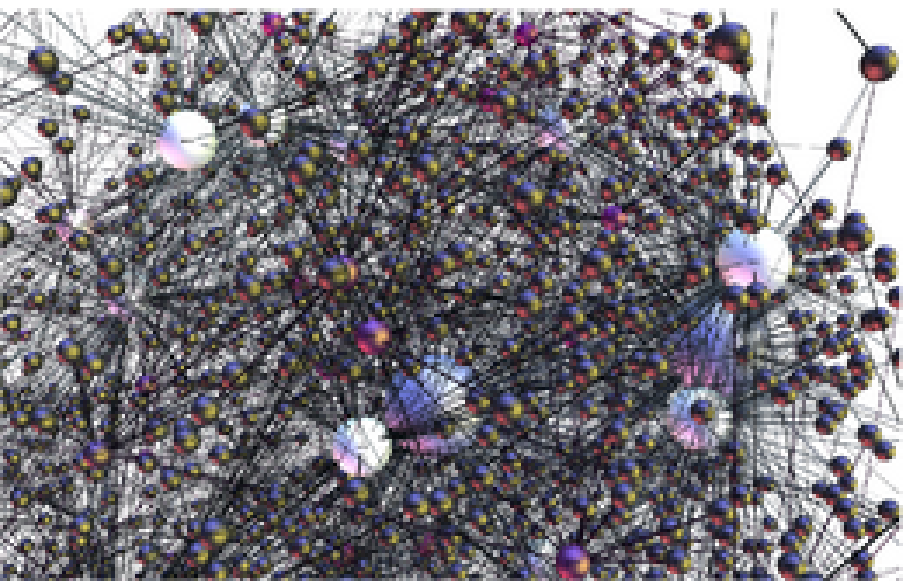}}\\
\end{tabular}
\caption{Left: Scatterplot of topological dimensions $dimQ^{i}$ of the leading
nodes in the layers where the propagated emotion opposes  the majority of
emotional messages: the nodes are sorted (along x-axis) according the $dimQ_{NP}^{i}$ in the positive layer when the majority of messages are
negative. Possition of the node along y-axis corresponds to its $dimQ_{PN}^{i}$ found in the negative layer when the majority of messages in
the network are positive. Right: Close-up of the part of the network in the
vicinity of big nodes (for the case of majority of positive messages).}
\label{fig-scatter_np_pn}
\end{figure}
The structure of emotion-propagating layers reveals altered position of individual nodes and their social capital. In Fig.\ \ref{fig-scatter_np_pn} we show the scatterplot of the topological dimension of each node in two counter-emotion layers. The color code indicates the number of nodes that correspond to a given combination of dimensions $(dimQ_{PN}^{i},dimQ_{NP}^{i})$. For the majority of nodes with at least one dimension larger than one, an off-diagonal pattern occurs. These findings indicate that the nodes have a different topological dimension, i.e. their neighborhoods differ in these two layers. A node that builds a compact neighborhood (large topological dimension) in the PN layer often has a small dimension (poor neighborhood) in the NP layer and vice versa. In other words, a large number of simplices, mostly $q=1$ simplices cf. Table\ \ref{tab-layers_q}, do not change their emotion polarity under the influence of the collective emotion in the network.

\section{Simmelian brokerage and the node's Q-vector \label{sec-brokerage}}

\subsection{Simmelian brokerage of nodes in online social networks\label{sec-Simmelian}}

In order to estimate the social capital of nodes (users) in the network, we
measure the Simmelian brokerage $B_{i}$ for each node $i=1,2,\cdots N_U$. According to Ref.\ \cite{panzarasa_brokerage}, for a given node $i$ Simmelian brokerage ``captures opportunities of brokerage between otherwise disconnected cohesive groups of contacts''. Quantitatively, $B_{i}$ is determined via the node's efficiency $E_{i}$ as \cite{panzarasa_brokerage}:
\begin{equation}
B_{i}=n_{i}-(n_{i}-1)E_{i}\ ,  \label{eq-brok}
\end{equation}
where  $n_{i}$ is the number of neighbors of the node $i$ inside a considered group $N_{i}$; the node's local efficiency $E_{i}$ is  determined by
\begin{equation}
E_{i}=\frac{1}{n_{i}(n_{i}-1)}\sum_{l\in N_{i}}\sum_{m\in N_{i}}
\frac{1}{d_{lm}}\ ,  \label{eq-efficiency}
\end{equation}
where $d_{lm}$ is the distance between all distinct  pairs of nodes $l\neq m$ in the set $N_{i}$ when  the node $i$ is removed. As stated earlier in our approach, the node's neighborhood $N_{i}$ is precisely defined at different topology levels by simplices and simplicial complexes in which the node $i$ resides.

\begin{figure}[!htb]
\centering
\begin{tabular}{ccc}
\resizebox{22.8pc}{!}{\includegraphics{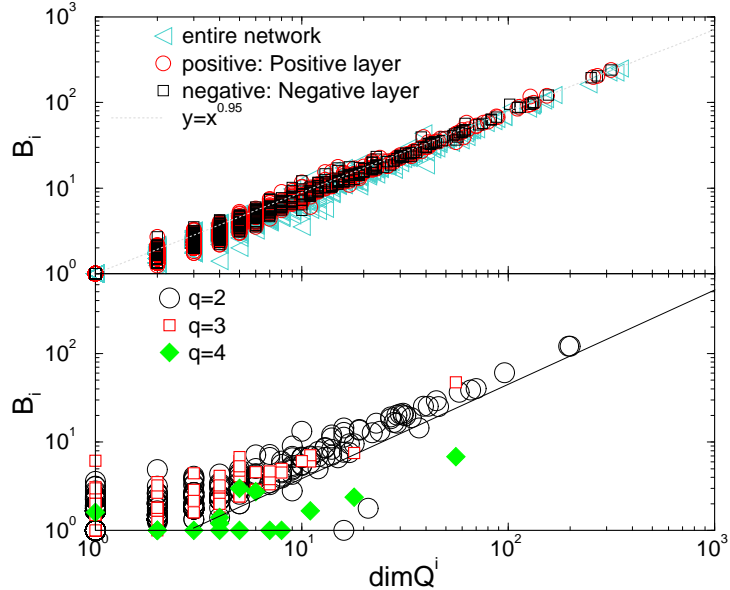}}\\
\end{tabular}
\caption{For the networks in Fig.\ \protect\ref{fig-osn_q}, Simmelian
brokerage $B_i$ of nodes  plotted against the node's topological dimension $dimQ^i$.  The symbols indicate social graphs at different topology levels $q=$2,3,4 (bottom panel), and the entire network $q=0$, and the two emotion-propagating layers (top panel).}
\label{fig-brkerage_Qi}
\end{figure}

Performing the computation indicated by Eqs.\ (\ref{eq-efficiency}-\ref{eq-brok}), we determine Simmelian brokerage $B_i$ for each node in the network shown in Fig.\ \ref{fig-osn_q}a.  In Fig.\ \ref{fig-brkerage_Qi}a  $B_{i}$ is plotted  against the node's topological dimension $dimQ^{i}$, where each point represents one node of the network. In addition, plotted are the node's brokerage values that are computed within the emotion-propagating layers.
The bottom panel includes similar plots but only for the higher topological levels $q=2,3,4$,   matching  the social graphs in  Fig. \ref{fig-osn_q}b,c,d,  respectively. Note that, in this notation,  the whole network corresponds to the level $q=0$. It is remarkable that, in these plots,  the majority of nodes follow a universal pattern that can be expressed by functional dependence
\begin{equation}
B_{i} \sim (dimQ^{i})^{\mu }\ ;  \label{eq-Bi-dimQi}
\end{equation}%
where the exponent $\mu \lesssim 1$ (see Table\ \ref{tab-nets}). Dispersion along vertical axis correlates with the number of higher-order cliques in the considered graph.

\begin{table}[!h]
\caption{For two influential nodes within the community in Fig.\ \protect\ref%
{fig-osn_q}, components of the topology vector $Q_i=%
\{Q^i_4,Q^i_3,Q^i_2,Q^i_1,Q^i_0\}$ and the related Simmelian brokerage are
computed in the whole network and in two emotion-propagating layers.
Different roles of the same node in the (counter-)emotion layer when the
majority of messages carries positive/negative emotion are demonstrated;
compared with the whole network indicates how the nodes build their social
capital by combining links at different layers.}
\label{tab-2nodes}\centering
\begin{tabular}{|c|c|c|c|c|}
\hline
$Node ID$ &  & $whole\ network$ & $NN$ & $NP$ \\ \hline
{1327} & $Q^i_q$ & \{3 53 144 46 0\} & \{- 1 18 83 0\} & \{- - 0 44 0\} \\
& $Brokerage$ & 162.79 & 95.44 & 43.85 \\
{2238} & $Q^i_q$ & \{0 0 96 253 0\} & \{- 0 0 316 0\} & \{- - 0 94 0\} \\
& $Brokerage$ & 250.80 & 239.69 & 90.49 \\ \hline
\end{tabular}%
\end{table}

\subsection{Relationship between the social capital and the topological dimension of nodes in a wider class of networks\label{sec-nets}}
In this section, we confirm the robustness of the functional dependence expressed by Eq.\ (\ref{eq-Bi-dimQi}). Particularly, we present an approximate analytical expression (which is exact in some limiting geometries) as well as numerical work for a wider class of networks.

Combining expressions (\ref{eq-brok}) and (\ref{eq-efficiency}), the brokerage of the node $i$ is given by
\begin{equation}
B_{i}=n_i-\frac{\sum_{l\in N_{i}}\sum_{m\in N_{i}}\frac{1}{d_{lm}}}{n_i}.
\label{brok1}
\end{equation}
where  the shortest paths between each pair of nodes $m\neq l$ within the graph are computed after the node $i$ is removed from that graph. This sum can be computed analytically in some limiting cases. For instance, consider the situation where the node $i$ belongs to a  $(q+1)$-clique, which  implies  that its number of neighbors is $n_i=q$ and all distances within the clique are $d_{lm}=1$. After removal of the node $i$, the remining nodes contribute to the sum as  $\sum_{l\in N_{i}}\sum_{m\in N_{i}}\frac{1}{d_{lm}} =q(q-1)$. A straightforward extension to the situation where the node $i$ connects $k$ such cliques leads  to $\sum_{l\in N_{i}}\sum_{m\in N_{i}}\frac{1}{d_{lm}} =kq(q-1)$. In this case, removal of the node leaves the cliques separated from each other, i.e., the distance between pairs of nodes from different cliques is infinite, while the distance inside each clique is one. Note that in the topology analysis described in sec.\ \ref{sec-topology},  the number of $(q+1)$ cliques related to the node $i$ is given by the node's $q$-level component, $Q_q^i$. Hence, the sum at $q$-level gives $\sum_{l\in N_{i}}\sum_{m\in N_{i}}\frac{1}{d_{lm}} =Q_q^iq(q-1)$. The situation is exact for top-level cliques, for example 5-cliques in Fig.\ \ref{fig-osn_q}d.

A similar reasoning can be extended to $(q-1)$-level, provided that at this level no shared  faces occur between the $q$-level cliques. Hence, in this case Eq.\ (\ref{brok1}) can be written as
\begin{equation}
B_{i}=\sum_{q}Q_{q}^{i}q-\frac{\sum_{q}Q_{q}^{i}q(q-1)}{\sum_{q}Q_{q}^{i}q} \ ,
\label{brok_e_1}
\end{equation}
where we also note that $n_i$ can be expressed via the components of the topology vector as $n_i= \sum_{q}Q_{q}^{i}q$.  It should be stressed that the expression is exact for those topology levels of a graph at which there are no shared faces of the higher level cliques, i.e, where the third structure vector $\hat{Q}_q =0$ (cf. Table\ \ref{tab-nets}). Otherwise, by extending  the summation over  topology levels, we obtain an approximate expression that can be written as
\begin{equation}
B_{i}=dimQ^{i}\langle q\rangle - \frac{\langle q^2\rangle}{\langle q\rangle} +1 \ ,
\label{brok_approx}
\end{equation}
 where we used the definition $dimQ^{i}=\sum_{q=0}^{q_{max}}Q_{q}^{i}$ and introduced an abbreviation $\langle q\rangle = {\sum_{q=0}^{q_{max}}Q_{q}^{i}q}/{\sum_{q=0}^{q_{max}}Q_{q}^{i}}$.

Note that in the case of tree structures, where the highest topology level corresponds to links, i.e., $q=1$ is the maximal clique, we have that $B_i=dimQ^i$ with the exact exponent $\mu=1$.  In a more complex network cliques of higher order occur and are weakly connected at lower topology levels, as in the case of studied online social network (OSN in Table\ \ref{tab-nets}). Consequently, the contribution of the second term in (\ref{brok_approx}) induces corrections eventually resulting with an exponent $\mu \lesssim 1$. The dispersion in the number of cliques in (\ref{brok_approx}) and the number of their shared faces at lower $q$-levels can be considerably greater in the case of more compact networks. Nevertheless, a power-law  dependence (\ref{eq-Bi-dimQi}) appears, with different values of the exponent $\mu$. The reasons for the occurrence of such power-law dependence in a general network structure  are not evident. Here, we provide a numerical proof by considering a wider class of networks.

\begin{figure}[!htb]
\centering
\begin{tabular}{cc}
\resizebox{16pc}{!}{\includegraphics{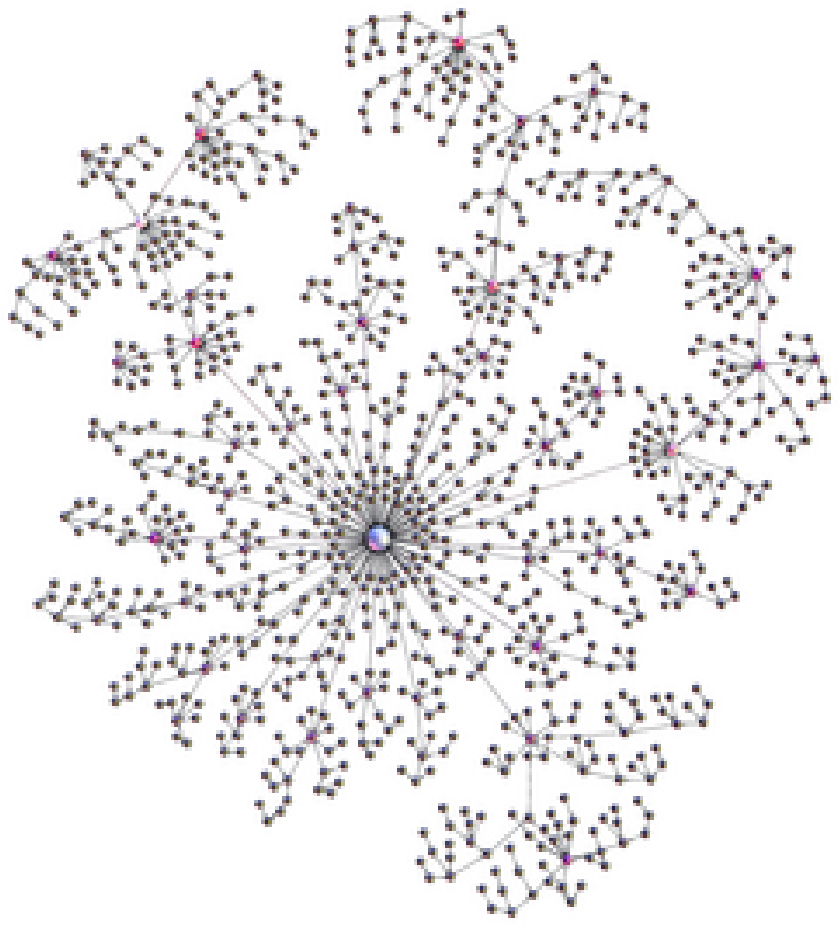}} &
\resizebox{16pc}{!}{\includegraphics{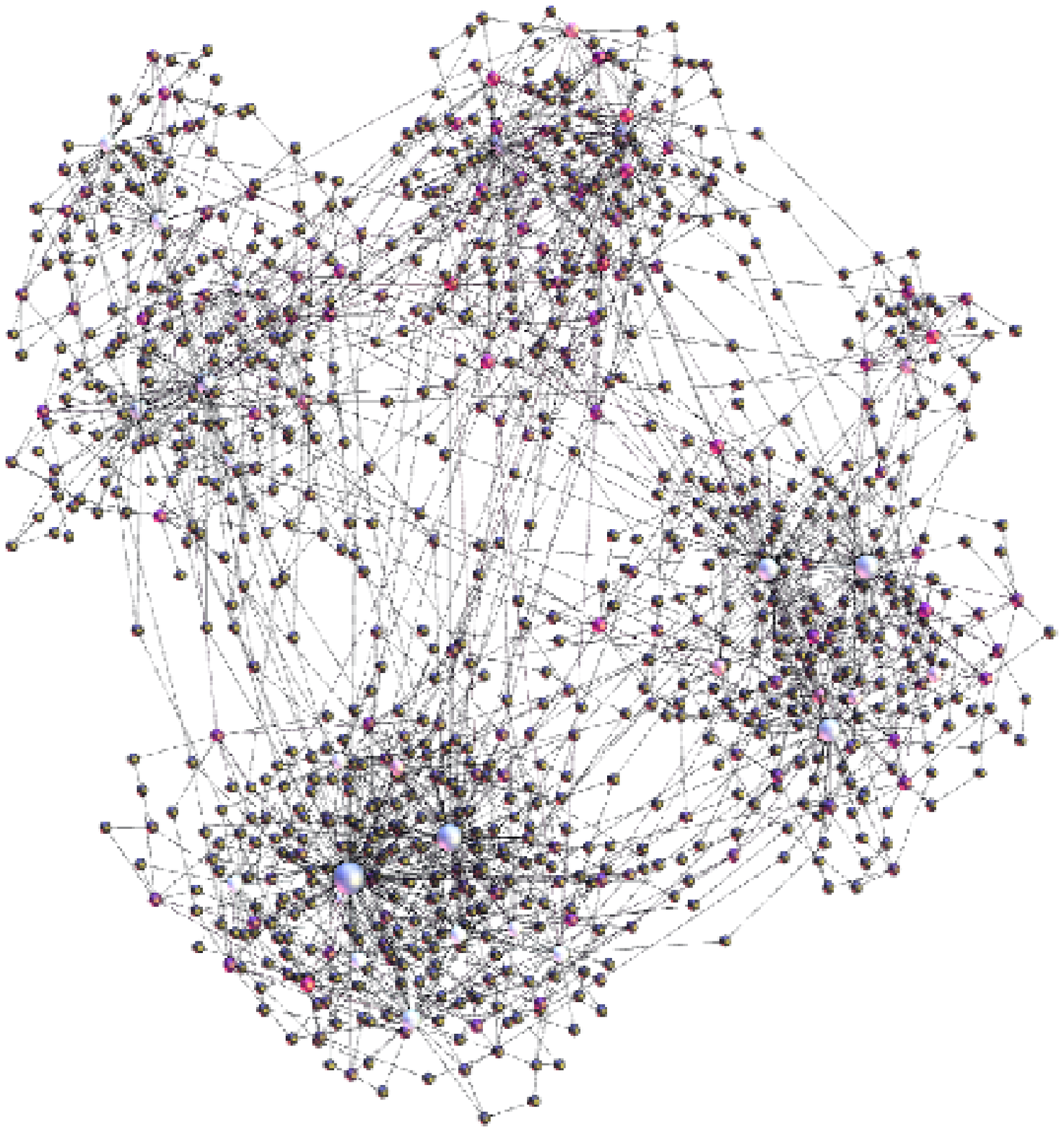}}\\
\resizebox{16pc}{!}{\includegraphics{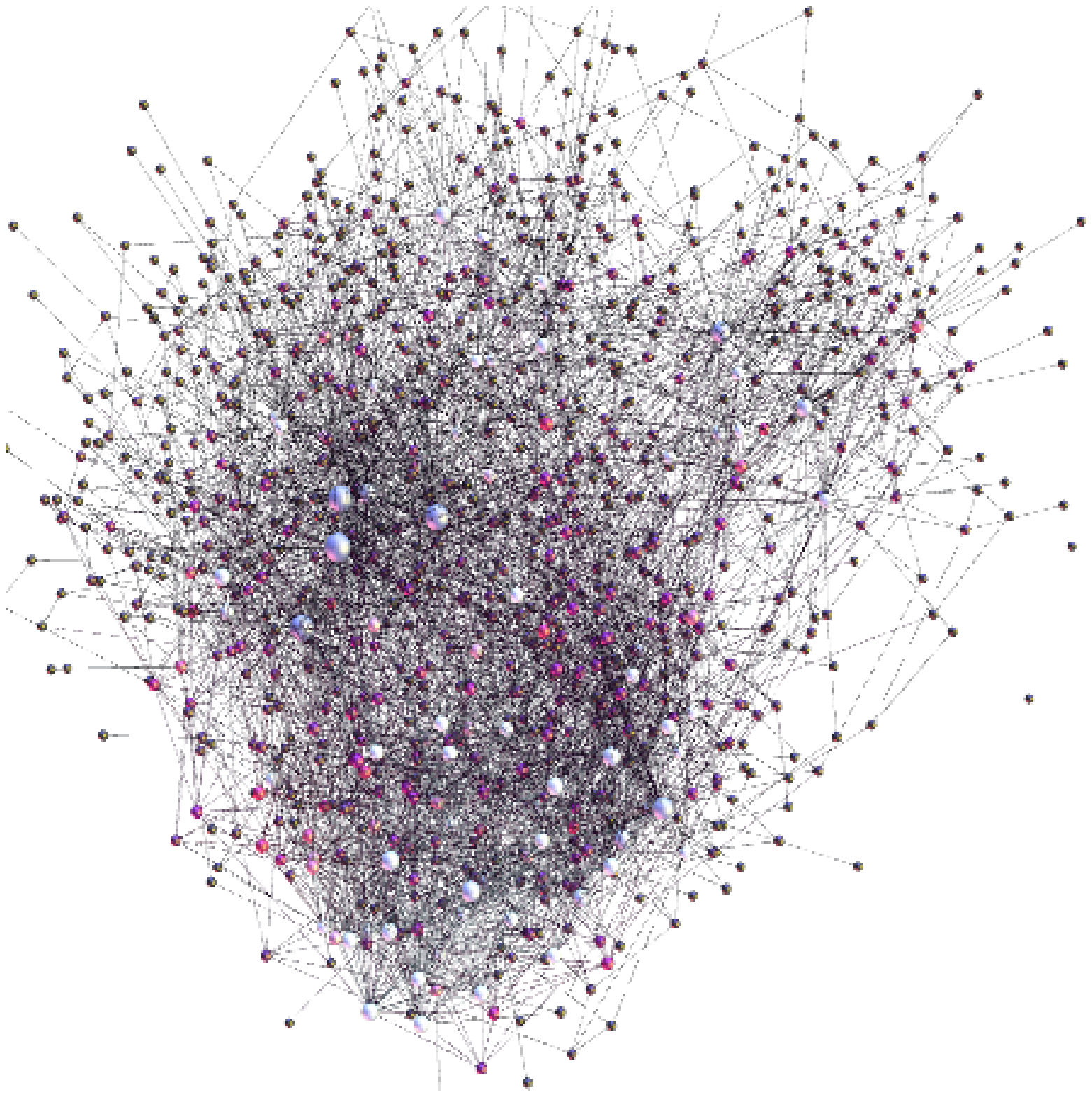}} &
\resizebox{16pc}{!}{\includegraphics{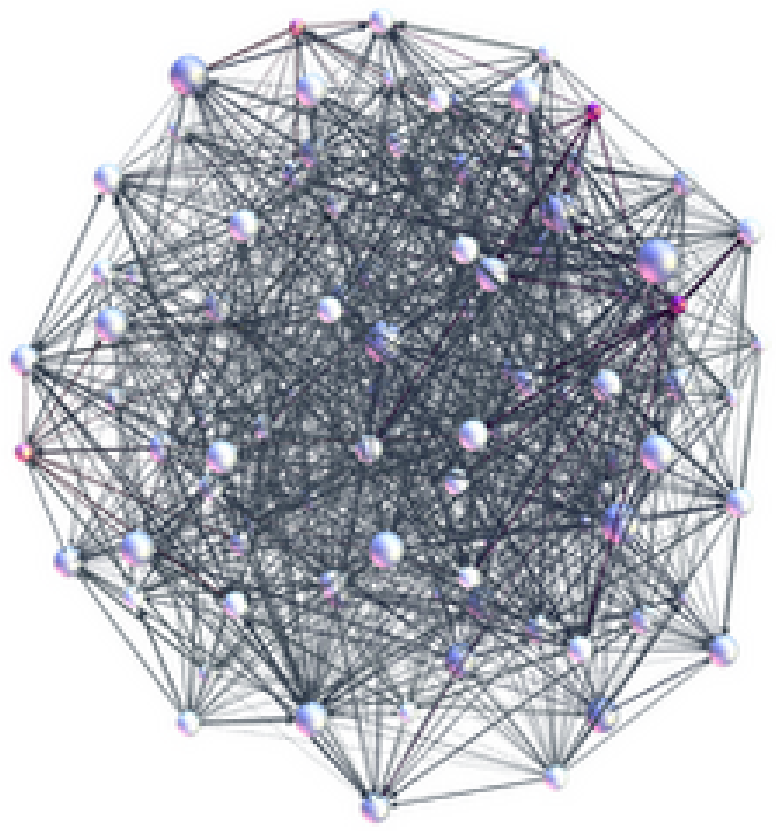}}\\
\end{tabular}
\caption{Computer-generated networks with different topology: (a) scale-free tree; (b)
network with clustered and weakly interlinked  scale-free communities; (c) same
parameters as (b) but stronger connections between communities; (d) a strongly
clustered single-community network.}
\label{fig-nets-grown}
\end{figure}

\begin{figure}[!htb]
\centering
\begin{tabular}{ccc}
\resizebox{22.4pc}{!}{\includegraphics{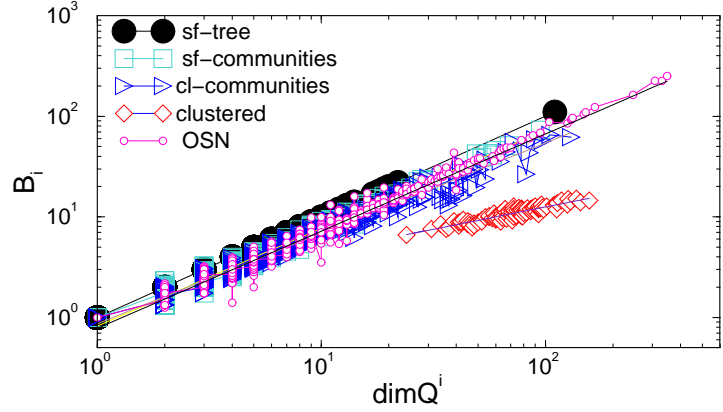}} \\
\end{tabular}
\caption{Brokerage versus topological dimension of nodes for
computer-generated network types and for OSN MySpace, as indicated. Network
characteristics are summarized in Table\ \protect\ref{tab-nets}.}
\label{fig-brkerage_Qi_nets}
\end{figure}

The analysis in previous sections suggests that the relation between Simmelian brokerage of nodes and their topological dimension depends on the graph architecture. Therefore, by varying the building rules of the network, one can vary the topological dimensions of different nodes and test the robustness of Eq.\ (\ref{eq-Bi-dimQi}). We consider several types that are shown in Fig.\ \ref{fig-nets-grown}. In these networks, the node's neighborhood can be varied by control parameters of the growth, ranging from the tree-like to a highly clustered compact structure.
These networks, consisting of approximately 1000 connected nodes, are generated by the algorithm that is initially described in Ref.\ \cite{mitrovic2009} for growth of scale-free networks with clustering and communities. The basic idea of clustered scale-free networks by preferential attachment and preferential rewiring of Ref.\ \cite{tadic2001} is implemented for the case where different communities (node groups) are allowed to grow. Thus, the attachment of new nodes is preferred within a currently growing community while rewiring can take part both within and outside of that community. Three parameters that control the structure are: $p$---the probability of a new community, $\alpha$ and $\beta$---that appear in the preferential shift-linear rules for  attachment and rewiring probabilities \cite{tadic2001}, respectively, and $M$---the number of nodes added per growth step. In addition, we consider a dense single-community graph consisting of 100 nodes, shown in Fig.\ \ref{fig-nets-grown}d.
The results of the topology analysis of these networks is summarized in Table\ \ref{tab-nets}. Topological dimension $dimQ^i$ of each node in these networks is also determined.
Then Simmelian brokerage is computed, according to the original formula in Eq.\ (\ref{brok1}), for each node and plotted against the node's topological dimension. The results are shown in Fig.\ \ref{fig-brkerage_Qi_nets}. The power-law dependence holds for each network type in the corresponding range of nodes topological dimensions. Values of the corresponding exponent $\mu$  are also given in Table\ \ref{tab-nets}.

\begin{table}[!h]
\caption{Extended topological properties of four computer-generated networks and OSN.}
\label{tab-nets}
\begin{tabular}{|l|lllll|cccc|c|}
\hline
network&$<k>$&CC& $d$& $<p>$& modul.& $q$ & $Q$ & $N_s$& $\hat{Q}$&$\mu$\cr
\hline
sf-tree&1 & 0& 20& 6.46& 0.916&  1& 998 & 998 & 0& \cr
       &  &  &   &     &      &        0 & 1& 998  & 0.99  &  1 \cr
\hline
sf-comm &1.94 &0.201 & 9& 4.42 &0.398 & 3  & 2   & 2  & 0& \cr
      &  &  &   &     &      &                        2  & 328 & 328& 0&  \cr
      &  &  &   &     &      &                   1 & 1322 & 1602&0.175&   \cr
&  &  &   &     &      &                        0 & 10  & 1611  &0.994&  0.974(5) \cr
\hline
cl-comm&3.83 &0.081 & 7 & 3.36&  0.395& 4 &16 & 16& 0&   \cr
      &  &  &   &     &      &        3 & 65  & 80  &0.188&   \cr
      &  &  &   &     &      &        2 & 837 & 900 & 0.07&  \cr
&  &  &   &     &      &        1 & 2269 & 3090  &0.269&   \cr
&  &  &   &     &      &        0 & 12  & 3101  &  0.996&  0.898(4)\cr
\hline
clustered& 14.52& 0.269&  2& 1.71& 0.207&  5 & 9& 9&0&  \cr
      &  &  &   &     &      &        4 & 264 & 266  & 0.008&  \cr
      &  &  &   &     &      &        3 & 1302  & 1522  &0.145&   \cr
&  &  &   &     &      &        2 & 367  & 1887  &0.806&   \cr
&  &  &   &     &      &        1 & 1 & 1887 &0.995&   \cr
&  &  &   &     &      &        0 & 1  & 1887 &0.995&0.44(2)  \cr
\hline\hline
OSN-&2.302 &0.183 &8 &4.067 &0.741 & 4 & 5& 5&  0& \cr
      &  &  &   &     &      &        3 &  90& 91  &0.011&   \cr
      &  &  &   &     &      &        2 & 1064 & 1103 & 0.035&  \cr
&  &  &   &     &      &        1 & 5397  & 6437  &0.161&   \cr
&  &  &   &     &      &        0 & 1  & 6437  &0.999&  0.969(2) \cr
\hline\hline
\end{tabular}
\end{table}

\section{Conclusions\label{sec-conclusions}}

The topological framework based on the simplicial complex representation of graphs offers systematic and in-depth characterization of complex  networks beyond the standard methods.  In this work, we have used simplicial complexes for addressing both local and global structures of online social networks. The studied graph  based on \texttt{MySpace} dialogs data from \cite{we-MySpace11}, is a prototypal online social structure: the network organization exhibits communities and layers closely related with the communication patterns between users and the dynamics of emotions.
The in-depth topology analysis reveals appearance of higher-order cliques and their complexes that are,  besides a tree-like local organization, often attached to some relevant nodes. Complementing the standard analysis of the network \cite{we-MySpace11}, the  three structure vectors introduced here, $Q$, $N_s$ and $\hat{Q}$, describe the network architecture at all topology levels from $q=0$ (the whole network) to $q_{max}$ (the level of maximal $(q_{max}+1)$-clique).  It demonstrates  that social network dynamics leads to unusual local structures, including cliques up to the fifth order and to complexes consisting of the cliques sharing triangle faces and tetrahedra.
The structure of emotion-propagating layers is much simpler, containing triangles as the highest order structures. It should be noted that, in another study \cite{ST_triadic}, triadic closure dynamics was recognized as one of the fundamental dynamic principles in social multiplex networks.

The node's $Q^i$-vector associated with every node of the simplicial complex,  defined here for the first time, differentiates in a precise manner different topological and hence graph  structures which contain the reference node. Its components $\{Q_q^i\}$  indicate the number of structures  at each hierarchical  level in which the node participates.  The total number of cliques of all sizes in which the node resides defines the node's topological dimension. The concept of the node's structure vector  has been useful in sequencing the social network, revealing a new insight into network organization from local to the global level and the role of individual nodes in it. In particular, ranking the nodes according to their topological dimension obeys  Zipf's law with two slopes. This feature helps to distinguish a smaller group of innovative nodes that build large surroundings or surroundings with a larger topological complexity, from the rest of the system.
Furthermore, we have demonstrated that this vector provides a useful quantitative measure of social capital of nodes in the communities and layers. Specifically, Simmelian brokerage, which measures the ability of a node to act as a broker between groups of other nodes in the recognized community or layer, scales as a power of the node's topological dimension. This functional relationship holds in a class of networks with varied composition. The scaling exponent has an exact value $\mu=1$ in tree structures and decreases towards lower values in the case of graphs that  contain a number of larger complexes (cliques with shared faces).

The influential nodes which possess considerable social capital can be further differentiated by considering the individual components $\{Q_q^i\}$ of their topology vectors.  Two types of  influential nodes can be observed. First, there are ``informer'' nodes that act as centers of star-like structures in the network; occurrence of such nodes is in the direct relationship with disassortativity of network's dynamical structure, observed in \cite{we-MySpace11}. Second, the nodes that
connect a large number of higher-order structures (for example in Fig.\ \ref{fig-osn_q}d) act as star centers in the \textit{conjugate} simplicial complex network. Their appearance may be related with  the disassortativity of the topological dimensions, demonstrated in the inset of Fig.\ \ref{fig-dimQ-Zipf}.  The influential nodes build their social capital by combining links in different layers. This conclusion is reached by inspection and comparison of the simplicial structure containing these nodes in the emotion-propagating layers   and the structure of the entire network, Table\ \ref{tab-2nodes}.

The metric based on the topological framework used in this work, can be also applied to discover ``key players''  in conventional social networks \cite{estrada2013social,nacher2014}. Similarly, it can be used to determine in-depth structure of techno-social networks that grow from scratch co-evolving  with the dynamics of chats and blogs \cite{we-entropy}, where the origin of communities and layers can be entirely different.  We expect  that this approach can be useful in a variety of other network-based studies of complex systems. Some examples are the brain dynamics and learning \cite{Mantzaris01062013}, and  innovation and collaboration systems \cite{katz2006,collaboration}, where different patterns of the actor's behavior are essential for the global dynamics.

\section*{Acknowledgment}
We thank for the support from program P1-0044 by the Research Agency of the Republic of Slovenia and from the European Community's COST Action TD1210 KNOWeSCAPE. S.M and M.R. would like to acknowledge the support from the Ministry of
Education and Science of the Republic of Serbia, under the project OI 174014. M.A. also wishes to thank for kind hospitality during his stay at the Department of Theoretical Physics, Jo\v zef Stefan Institute, where this work was done.


\end{document}